\documentstyle[11pt,newpasp,twoside,epsf]{article}
\def\edcomment#1{\iffalse\marginpar{\raggedright\sl#1\/}\else\relax\fi}
\reversemarginpar

\newcommand{\fig}[6]{
    \protect\centerline{
    \epsfxsize=#1\epsffile[#2 #3 #4 #5]{#6}
    }}

\begin{document}

\title{The Frequency of Active and Quiescent Galaxies with Companions}
\author{Henrique R. Schmitt}
\affil{National Radio Astronomy Observatory, 1003 Lopezville, Socorro, NM87801, USA}

\setcounter{page}{1}
\index{Schmitt, H. R.}

\begin{abstract}
We study the percentage of active, HII and quiescent galaxies with companions
in the Palomar survey. We find that when we separate the galaxies by their
morphological types (ellipticals or spirals), to avoid morphology-density
effects, there is no difference in the percentage of galaxies with companions
among the different activity types.
\end{abstract}

\section{Introduction}
One of the major concerns in the study of AGNs is the mechanisms responsible
for making gas lose angular momentum and move from galactic scales down to the
inner $\sim$1 pc region of the galaxy. Several mechanisms have been suggested
to explain how this is possible, such as interactions or bars
(see Combes contribution to this proceedings for a review).

Several papers addressed the influence of interactions on the fueling of
AGNs, however, there is no concensus on the subject. Some studies suggest
that there is a larger percentage of companions around Seyferts (Dahari 1984),
while others do not find a significant difference (Fuentes-Williams \&
Stocke 1988). In contrast, in the case of luminous starburst galaxies it is
well known that they are related to interacting systems (Sanders et al. 1988).

We believe that most of the problems in this field are due to the way
the AGN and control samples are selected. A proper comparison 
between AGNs and normal galaxies requires the samples to be selected by
a matching host galaxy property. We solve this problem by using the
Palomar sample (Ho et al. 1997), which contains homogeneous classification
of all galaxies brighter than B$=12.5$ in the northern hemisphere.

\section{Results}
The detailed results of this study were published in Schmitt (2001), and our
principal results are presented in Figure 1. We consider that a galaxy
has a companion if the distance between it and the possible companion
is smaller than 5 times the diameter of the primary galaxy. Also, the
brightnesses of the galaxies cannot differ by more than 3 magnitudes,
and the difference between their radial velocities should be smaller
than 1000 km~s$^{-1}$.

\begin{figure}[t]
    \fig{12cm}{20}{150}{590}{430}{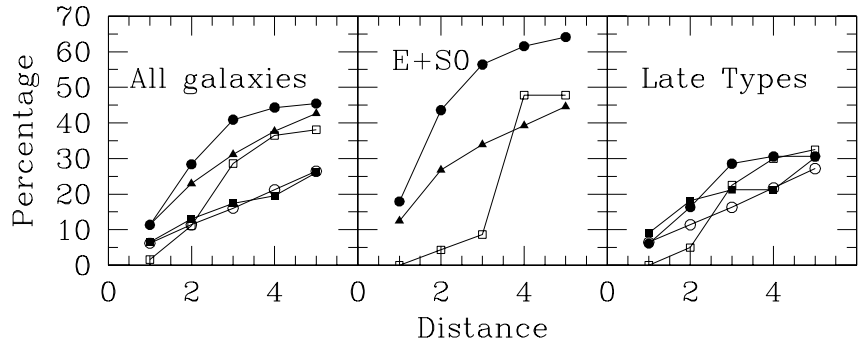}
\caption{Percentage of galaxies with companions as a function of the distance
between the two galaxies, in units of the diameter of the primary galaxy.
LINER's are represented by filled circles, Transition galaxies by open
squares, Absorption line galaxies by filled
triangles, Seyfert galaxies by filled squares and HII galaxies by open
circles. The left panel shows the results for all the galaxies, the middle
panel shows the results for Elliptical and S0 galaxies only, while the
right panel shows the results for late type galaxies only.}
\end{figure}

The left panel of Figure 1 shows the percentage of galaxies with companions
as a function of the distance between them for the 5 activity classes
defined by Ho et al. (1997). Contrary to previous results,
we found that LINER's, Absorption line and Transition galaxies
have a higher percentage of companions when compared to Seyfert and HII
galaxies. However, we did not take into account the morphological types
of the galaxies. A large portion of LINER's, and transition galaxies,
as well as almost all Absorption line objects, are in Ellipticals or S0's,
while a considerable portion of Seyferts and HII galaxies are found in
late type galaxies. When we consider only Ellipticals and S0's (middle
panel in Figure 1), or only galaxies with morphological type Sa and later
(right panel), we see that there is no difference in the percentage of
galaxies with companions for different activity types, indicating
that the above result was due to the morphology-density effect. Implications
of this result are discussed in Schmitt (2001).

\acknowledgements
The National Radio Astronomy Observatory is a
facility of the National Science Foundation operated under cooperative
agreement by Associated Universities, Inc.

\end{document}